\documentclass[twocolumn,showpacs,preprintnumbers,amsmath,amssymb,prb,aps]{revtex4-2} 
\usepackage{epsfig}
\usepackage{epstopdf}
\usepackage{multirow}
\usepackage{graphicx}
\usepackage{dcolumn}
\usepackage{bm}
\usepackage{textcomp}
\usepackage{array}
\usepackage{romannum}
\usepackage{csquotes}
\usepackage{subcaption}
\usepackage{booktabs}
\usepackage{amsmath}
\usepackage{mathtools}
\usepackage{enumitem}
\usepackage{hyperref}
\hypersetup{
    colorlinks=true,
    linkcolor=blue,
    filecolor=magenta,      
    urlcolor=blue,
    pdftitle={Rahul Raj},
    }
    
\begin{document}
\title {Studying the Seebeck coefficient and exploring the possibility of enhancing ZT upto 1.8 for NaCo$_2$O$_4$ in high temperature region}
\author{Rahul Raj}
\altaffiliation{ \url{rahulraj0301@gmail.com}}
\author{Sudhir K. Pandey}
\altaffiliation{ \url{sudhir@iitmandi.ac.in}}
\affiliation{School of Mechanical and Materials Engineering, Indian Institute of Technology Mandi, Kamand - 175075, India}

\date{\today}

\begin{abstract}
Here, we have studied the temperature dependent Seebeck coefficient (S) of the NaCo$_2$O$_4$ (NCO) by using experimental and computational methods. The range of experimentally obtained S is $\sim$55 to 103 $\mu$V/K in the temperature range of 300-600 K, which confirms the p-type behaviour of NCO. The electronic structure of this compound is obtained via DFT+\textit{U} formalism. The band dispersion and partial density of states confirms the magnetic and half metallic nature. Furthermore, in the transport properties, the obtained S using a \textit{U} = 4 eV gives the best match with experimental data. The temperature and chemical potential dependent S$^{2}$$\sigma$$/$$\tau$ is calculated using the obtained electronic transport properties, in which the maximum value obtained for p(n)-type doping is $\sim$22(61)$\times$10$^1$$^4$ $\mu$WK$^{-2}$cm$^{-1}$s$^{-1}$. The possibility of enhancing the ZT is identified, and it is calculated in temperature range 300-1200 K. The maximum calculated value of ZT is 0.64 for p-type and 1.8 for n-type doping at 1200 K. The calculated carrier concentration obtained for p(n) type doping at 1200 K is $\sim$1.17 (1.6)$\times$10$^2$$^2$ cm$^-$$^3$. This study suggests that the careful doping of p and n type can enhance the applicability of this compound in thermoelectric for high temperature application.
 
\end{abstract}

\maketitle

\subsection{Introduction} 
Nonrenewable energy sources are limited in nature, and the energy requirements are increasing exponentially with time \cite{panwar,lior}. Therefore, moving towards renewable energy became necessary for the fulfillment of required energy demand. We have some renewable energy sources which can assist non-renewable energy to fulfill the requirement, they are like solar energy, hydro-power, bio-gas, wind energy, hydrogen, thermoelectric (TE), etc \cite{panwar,lior}. Among these energy sources, TE has been proving itself as a potential candidate due to the direct conversion of waste heat into electricity \cite{mamur,jaziri}. The operation of TE devices without any waste products and with large durability are the major advantages \cite{jaziri,date}. The efficiency of these TE devices depends mainly on power factor (PF) and the figure of merit (ZT). Both can be expressed as follows \cite{jaziri}:
\begin{equation}
  PF={S ^2\sigma}
\end{equation}
Where S, and $\sigma$ are the Seebeck coefficient (S) and electrical conductivity ($\sigma$), respectively.
\begin{equation}
  ZT=\frac{S ^2\sigma T}{\kappa}
\end{equation}
Where $\kappa$ is thermal conductivity due to electrons ($\kappa_e$) and phonons ($\kappa_{ph}$), respectively, and T is an absolute temperature.

The ZT should be higher ($\geq$1) \cite{snyder} for the good TE materials. In order to achieve higher ZT, the materials like perovskites, lead and bismuth chalcogenides, silicon-germanium based alloys, metal oxides, Clathrates, half-Heusler and full-Heuslar alloys, organic materials, etc, have been explored \cite{date}. Till now, the best commercially available TE materials are Bismuth (Lead) based chalcogenides, which achieve maximum ZT upto $\sim$1 or beyond, around room temperature and above \cite{witting,zhang2012,lian,mehta2012new}. However, these materials have limitations of thermal and chemical stability in the higher temperature. In order to overcome these limitations, the metal oxides can be one of the best options in TE due to its various advantages, like high S, more stability, oxidation free, etc. in a higher temperature range, being non toxic in nature and having a simple synthesis process. 
In 1997, Tersaki \textit{et al}. reported a new layered metal oxide, the single crystal NaCo$_2$O$_4$ (NCO), with the high S (100 $\mu$V/K) at 300 K \cite{terasa}. NCO is metallic in nature, and the high value of S was surprising for the metallic compound. Thus, this study opened the door for researchers to take an interest in the experimental study of S. 

In view of the high S, extensive studies have been done on the Na$_x$Co$_y$O$_2$ by different groups, including various substitutions on the place of Na and Co through experimental and computational approaches in a wide temperature range \cite{terasa,singh,kurosaki,ito,fujita,ito2,seetawan2,altin,park,katsuyama,gao,ma,assadi2019,xu,cheng,terasaki2001,liu2,ermawan,tanaka,terasaki1998,okane,kitawaki,erdal}. 
In these works, the centre of study was NCO, which shows the wide range of S across various temperature domains.  
The values of S for NCO at 300 K were reported as $\sim$41, $\sim$80 and $\sim$80 $\mu$V/K by Ermawan \textit{et al}., Kitawaki \textit{et al}. and Cheng \textit{et al}., respectively \cite{ermawan,kitawaki,cheng}. Further, the S values were found as $\sim$57 $\&$ 69 $\mu$V/K in the work of Erdal \textit{et al}. at 300 K \cite{erdal}. In these studies, the different synthesis processes were used, like solid state reaction (SSR) technique, sol-gel method, cold pressure compacting of powder sample, electrospinning, etc. These synthesis processes may lead to the variation in the value of S. Identifying the variation in S, it can be said that the approach to achieve the unique value of S at a given temperature is still a complex problem, which gives direction to further exploration of the S through an experimental approach. 

From the above reported work, it is said that the comprehensive experimental studies have been carried out. However, very few computational studies for the Na and Co based oxides are reported for the electronic structure \cite{singh2007,xu,ma,assadi2012plane,assadi2019} and electronic transport properties \cite{assadi2019,gao,kuroki2007pudding,xiang}. In the electronic structure, the partial density of states (PDOS) and band dispersion curve are analysed, while in electronic transport properties, mainly the S is calculated. At 300 K, the calculated values of S by Gao \textit{et al}. were 38 and 22 $\mu$V/K within the DFT based virtual crystal approximation and rigid band model, respectively, for NCO \cite{gao}. These values are inconsistent with the experimental works of S reported in Refs. \cite{tanaka, kitawaki, kuroki2007pudding, xiang}. In these studies the strong correlation effect was not considered, that is the case here due to the presence of partially filled Co 3\textit{d} orbitals. In the work of Assadi \textit{et al}., the calculation of S was done for the doped element as Gd and Yb in NCO by using Koshibae's equation \cite{koshibae2000thermopower} and gets the value of 154 $\mu$V/K \cite{assadi2019}. In the equation, the calculation of S was performed by considering the carrier density of Co ions and possible electron arrangements, which may not be appropriate for the variation of S with temperature. Further, the S was calculated by Kuroki \textit{et al}. for Na$_x$CoO$_2$ by using Boltzmann’s equation, and obtained the high values (100 to 250 $\mu$V/K) by varying \textit{x} \cite{kuroki2007pudding}. This calculation suggested that the high S came from the pudding mold bands, which originates from the orbitals degeneracy. It was calculated only upto 300 K. In addition to these, Singh had calculated S on the basis of LDA band structure by using kinetic transport theory model and got the value of 110 $\mu$V/K at 300 K \cite{singh} for NCO. However, he did not calculate the S in the high temperature range by considering strong correlation effect. On the top of the above, Shamim \textit{et al} have calculated the S for the temperature beyond 300 K and also calculated ZT with constant relaxation time for the Na$_0.$$_7$$_4$CoO$_2$ \cite{sk2020}, but not for NCO. From the above calculated data of the Na and Co based oxides, it can be said that the experimental analysis of these oxides through computational approach by considering strong correlation effect in wide temperature range is still unexplored. Also, the proper value of onsite coulomb interaction for Co 3\text{d} orbitals are unanswered. These limitations provide an opportunity for the calculation of S by considering the strong correlation effect in the wide temperature ranges, which can explain the experimental S. Therefore, we have taken an attempt to explore the experimental S by computational approach after considering strong correlation effect for the temperature range 300-600 K.

The high S obtained from the experimental and computational approach for NCO, compelled the researchers to pay attention towards the TE applications of this material. In this direction, the study of ZT has been performed by various groups, in which the high S and other transport properties ($\sigma$ and $\kappa$) have been taken into account \cite{kurosaki, ito, zhang2019, perac}. The value of ZT reported by Kurosaki \textit{et al}. was in the range of $\sim$0.005-0.019 for the temperature domain $\sim$324 to 730 K, where the sample was prepared by hot pressing \cite{kurosaki}. In the work of Ito \textit{et al}., the ZT was increased from $\sim$0.15 (at 464 K) to  0.5 (955 K), in which SSR method is used for the synthesis \cite{ito}. Further, the reported value of ZT by Zhang \textit{et al}. is $\sim$0.0024 and $\sim$0.06 corresponding to $\sim$400 and $\sim$930 K, respectively \cite{zhang2019} by SSR method. These variations in ZT indicate that its optimisation is still an ongoing process for NCO, which provides a room for the researchers to think in the direction of enhancement and optimisation of ZT. In this context, changing the carrier concentration by doping \cite{af1957} and structural changes by nanostructuring \cite{ma201,solet2025band} can be the better option to get high PF and low $\kappa$. Thus, many studies have been carried out by different groups with various dopants to enhance S and other transport properties   \cite{assadi2019,altin,ito,ito2,seetawan2,fujita,park,kurosaki,zhang2019, perac}. 
Altin \textit{et al}. have reported the S value in the range of $\sim$60-90 $\mu$V/K at 300 K for Boron doped Na$_0.$$_7$Co$_1$$_-$$_x$B$_x$O$_2$ (x = 0-0.75) \cite{altin}. In this work, the value of ZT is found to be decreased after doping. In the Ag doped compound, Na$_x$Co$_2$$_-$$_y$Ag$_y$O$_4$ (x$\sim$1.5 Y = 0-0.5), the obtained value of S is $\sim$120–170 $\mu$V/K at 1000 K corresponding to different y, in which the ZT is maximum for y = 0.2 \cite{seetawan2}. In this study the maximum value of ZT is found to be 0.124 at 973 K. The maximum ZT reported by Ito \textit{et al}. was 0.8 for Na$_1$$_.$$_7$Co$_2$O$_4$, in which the decrement in $\kappa$ was observed for $\sim$450 to $\sim$750 K and then increases upto $\sim$950 K \cite{ito}. Fujita \textit{et al}. had reported the value of the S as 200 (170) for single crystal (polycrystal) with a maximum ZT of 1.2 for Na$_x$CoO$_2$$_-$$_\delta$ single crystal at 800 K\cite{fujita}.
Further, Klyndyuk \textit{et al}. have obtained the ZT value in the range of 0.05-0.22 in the study of Na$_0.$$_8$$_9$Co$_0.$$_9$M$_0.$$_1$O$_2$ (M = Ni, Cu,Mo, Bi) \cite{klyndyuk}. In addition to these, Shamim \textit{et al}. have proposed the ZT$>$2 by computational approach for Na$_0.$$_7$$_4$CoO$_2$ \cite{sk2020}.  
Thus, from the above achieved and proposed value of ZT, we have observed that it is found changing with the doping element and quantity. The observed changes are an indicator of the possibility of enhancement in ZT. 
To take advantage of this possibility, we have calculated PF and corresponding ZT by using calculated S, $\sigma$ and reported $\kappa$ \cite{ito}.

In this work, first we have studied the S of NCO by using combined experimental and computational approach. The monotonically increasing nature of S is found, in which the obtained value is $\sim$55 (103) $\mu$V/K at 300 (600) K. The calculation of electronic transport properties in DFT+\textit{U} (\textit{U} = 4 eV) are found to explain the experimental S in a better way. The temperature and chemical potential ($\mu$) dependent S$^{2}$$\sigma$$/$$\tau$ has been calculated by using the electronic transport properties. By studying S$^{2}$$\sigma$$/$$\tau$, the possibility of enhancing the ZT upto 0.64 (1.8) for p(n) type doping at 1200 K is identified. The calculated carrier concentration is 1.13$\times$10$^2$$^2$ cm$^-$$^3$ at 100 K, and at 1200 K it is found to be 1.17$\times$10$^2$$^2$ (1.6$\times$10$^2$$^2$) cm$^-$$^3$ for p(n) type of doping. 


\subsection{Experimental and Computational Details} 
We have synthesized the compound by solid state reaction method \cite{cheng,terasaki2001,kurosaki}. Na$_2$CO$_3$ and Co$_3$O$_4$ have been taken as base material in the required stoichiometric ratio. We have baked the powder for 2 hrs at 120$^\circ$C to remove the possible moisture. The baked powders have been mixed properly by using a mortar and pestle for $\sim$7 hrs followed by compacting into pellets and then subjected to heat treatment at 1133 K for 10 hrs in an open-air furnace. The sintered pellets were pulverised into powder, then compacted into pellets and again given heat treatment at 1173 K for 10 hrs. 
This pellet is used for the measurement of the S by using home made instrument \cite{sk2022i}.

In order to explain the experimental results of NCO, density functional theory (DFT) \cite{kohn} and DFT+U calculations for electronic structures and electronic transport properties have been performed by using LDA exchange correlation functional in ABINIT software package\cite{gonze}. 
The cut-off energy (E$_{cut}$) has been taken as 40 Ha along with a Monkhorst-pack grid of 32$\times$32$\times$8. We have taken the tolerance on difference in total energy as 10$^-$$^6$ eV. By considering the above parameters, the calculation for transport properties has been performed for various onsite coulomb interaction (\textit{U} = 2, 3, 4 and 5 eV), in which \textit{U} = 4 eV is found suitable for the explanation of experimental result in a better way. The S and $\sigma$ are calculated by using the BoltzTraP tool \cite{madsen}. 
\section {Result and discussion}
\subsection {Structural characterization}
The room temperature X-ray diffraction (XRD) of the powder sample has been carried out for the structural characterization of the sample.
The data obtained from the XRD has been used for the fitting through the Rietveld refinement. In order to fit NCO peaks, the initial structural parameters have been taken from Jansen \textit{et al}. \cite{jansen}. Fitting of the data reveals the absence of Bragg's positions at 2$\theta$ of $\sim$31.5$^\circ$, $\sim$59.5$^\circ$ and $\sim$64.5$^\circ$, which confirms the presence of another phase other than NCO in the compound. The existence of Co$_3$O$_4$ at the same positions \cite{ma201,okane} has been confirmed by the literature. Therefore, we have performed the dual phase Rietveld refinement. In order to fit the Co$_3$O$_4$ peaks, the structural parameters have been taken from the Cardenas-Flechas \textit{et al}. \cite{cardenas}. The convergence factor ($\chi$$^2$) value obtained from the refinement is 2.55, which indicates the goodness of the fitting. The fitting of the data confirms the primary phase of NCO ($\sim$90$\%$) along with a minor phase of Co$_3$O$_4$ ($\sim$10$\%$) in the compound. Fig.1 shows the refined XRD pattern of the compound.  
In the figure, the black, red, green, blue and violet colours represent the observed, calculated, difference between observed and calculated, Bragg's position and positions of Co$_3$O$_4$,  respectively.
The Hexagonal crystal structure with space group P6$_3$22 (182) of the compound is confirmed by the Rietveld refinement. The lattice parameters obtained from refinement are a = b = 2.82 $\AA$ and c = 10.94 $\AA$. These parameters are in good agreement with the previously reported lattice parameters \cite{liu2,singh,tanaka}. Moreover, the crystal structure of NCO have been generated in the Vesta software package by using the structural parameters obtained from the refinement. Fig.2 shows the Hexagonal crystal structure of the compound. The figure demonstrates the layered structure of NCO, which consists of the Na layer in an alternative manner with the Co-plane. The unit cell of NCO is formed by distorted CoO$_6$ octahedrons, where Co located at the centre of the octahedrons while the equatorial and axial positions are occupied by six O atoms. The bond length and bond angle between respective elements are shown in Table 1.
\begin{table}
\begin{center}
\caption{\label{tab:table1}%
\small{Bond length and angle for Hexagonal crystal structure of NaCo$_2$O$_4$}}
\setlength{\tabcolsep}{9pt}
\begin{tabular}{lcc}
\textrm{Type of bond}& 
\textrm{Experimental bond length($\AA$)}\\ 
\colrule 

\colrule

\colrule

Co-O  &  1.86\\
Na-O  &  2.45\\
Na-Co & 2.73\\
\colrule

\colrule
\midrule
\textrm{Bond type}&
\textrm{Experimental bond angle}($^\circ$C)\\ 
\colrule 
O(1)-Co-O(1) &  98.73\\
O(2)-Co-O(2) &  81.26\\
O(1)-Na-O(1) & 70.14\\
O(2)-Na-O(2) & 98.86\\

\colrule
\end{tabular}
\end{center}
\end{table}

\begin{figure}
     \includegraphics[width=0.9\linewidth]{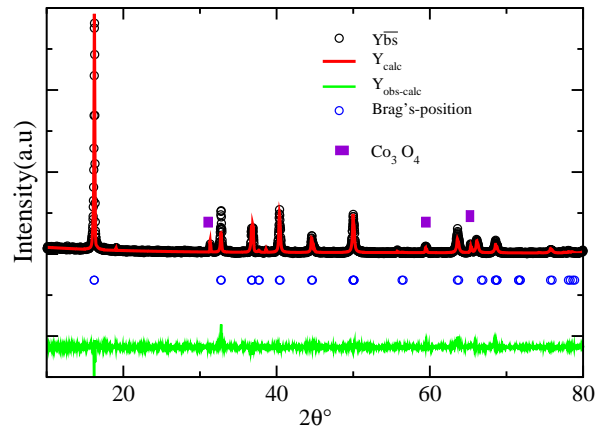}
     \caption{The powder XRD pattern of synthesized sample}
     \label{fig:my_label}
\end{figure}
\begin{figure}
     \includegraphics[width=0.7\linewidth,height=0.8\linewidth]{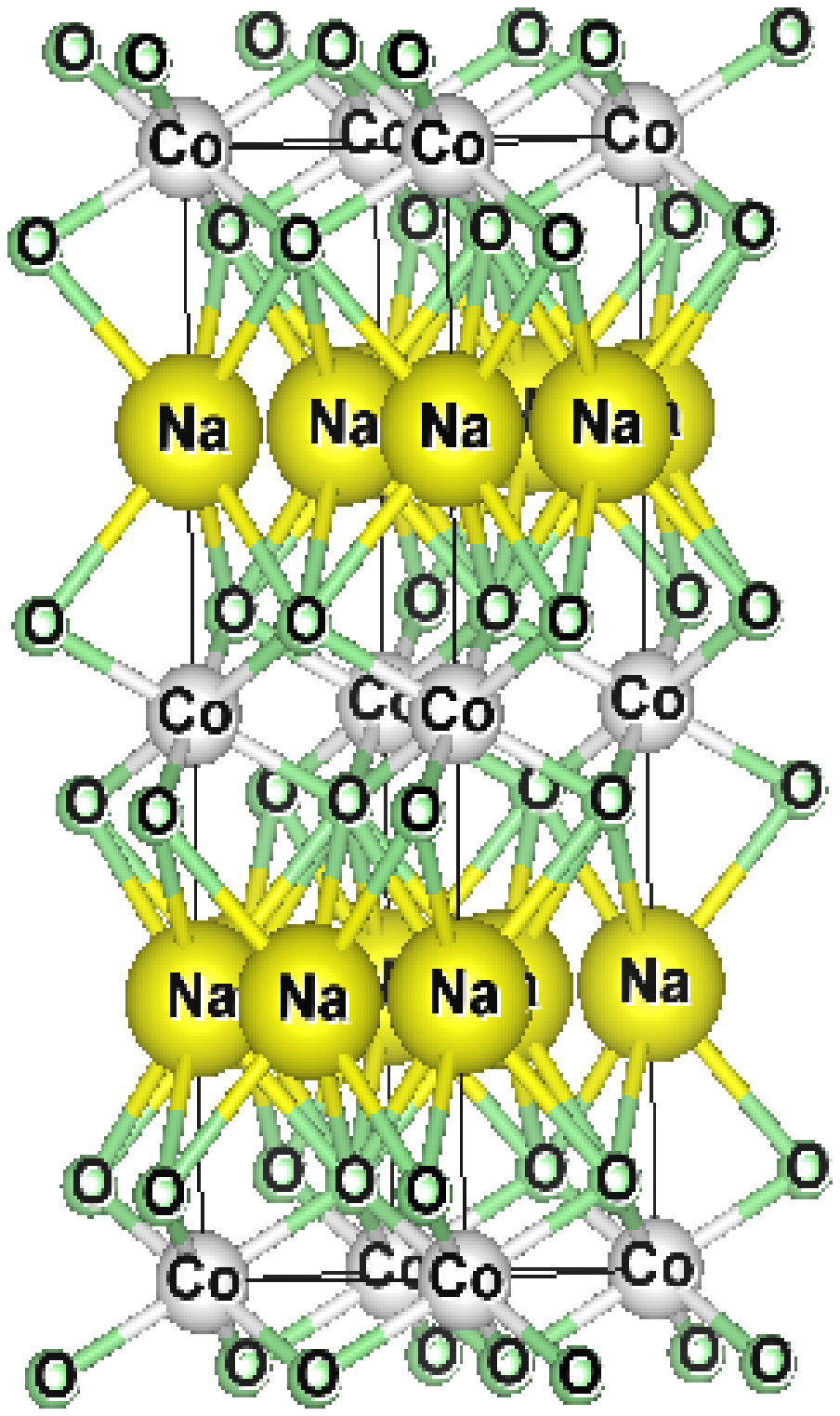}
     \caption{Hexagonal Crystal structure of NaCo$_2$O$_4$}
     \label{fig:my_label}
\end{figure}

\subsection{Seebeck coefficient}
 The charges in the solid materials start moving in a particular direction, when subjected to the thermal gradient. These movements of charges lead to non-zero charge distribution inside the solid, which generates a macroscopic TE field, which is called Seebeck effect. This effect can be measured as thermovoltage (V$_T$) with respect to temperature difference ($\Delta$T) \cite{sehnem}, called Seebeck coefficient. It is well known that the material's performance in TE is governed by its figure of merit in which the parameter S plays a vital role. In order to investigate the TE performance of NCO, the S has been studied.

We have measured the S of the compound in the temperature range 300-600 K.
We have seen in the structural analysis, that there is a presence of a small amount of Co$_3$O$_4$ in the compound. Therefore, it might be questionable that these S can be affected by the presence of Co$_3$O$_4$. However, due to large band gap of Co$_3$O$_4$ (1.6-2.1 eV) \cite{bhargava2018investigation} in the given temperature range, the contribution in the S can be neglected. Therefore, the S obtained from the measurement is mainly contributed by NCO. Fig.3 shows the experimentally measured S.
The value of S is found to be positive in the studied temperature range. The positive value of S in the overall temperature range confirms the dominating p-type behavior of the compound. This S is also increasing monotonically in the given temperature range, and the value at 300 K is found to be $\sim$55 $\mu$V/K, while at 450 (600) K it is $\sim$86 (104) $\mu$V/K. 
This monotonic increasing nature of S is not seen in the many standard TE compounds like in n-type Bi$_2$Te$_3$. In Bi$_2$Te$_3$, the value of S increases upto $\sim$400 K, after which it starts decreasing \cite{yang}.
Therefore, NCO is considered as a promising compound for the high temperature TE applications due to the increasing trend of S at elevated temperature. 

\begin{figure}
\includegraphics[width=8.5cm, height=6.5cm]{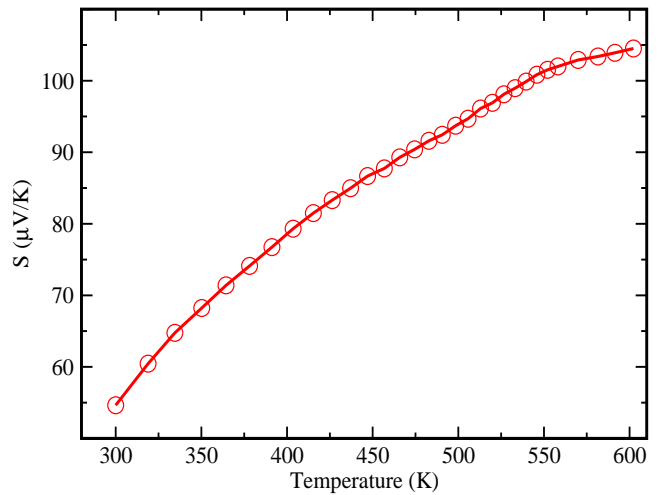} 
\caption{Variation of experimental Seebeck coefficient of NaCo$_2$O$_4$ with temperature.} 
\end{figure}

\subsection {Partial density of states} 
PDOS have been calculated for NCO to get the idea about the orbital's contribution of the elements in the electronic, transport and magnetic properties, respectively. We have performed spin unpolarized (SUP) and spin polarized (SP) calculations using the DFT. In addition to this, the SP calculation in DFT+\textit{U} has been also calculated. 
Fig.4 represents the SUP and SP solutions within the DFT and SP solution in DFT+\textit{U}. The dashed line at 0 eV corresponds to the Fermi level (E$_F$). 

Fig.4(a) represents the SUP solution obtained from the DFT. In order to analyse different aspects of PDOS, the figure is divided into the three regions. Where, region I, region II and region III are taken from -7 to -1.7, -1.7 to 0.5, and 0.5 to 2.5 eV, respectively. In the region I, the contribution from Co 3\textit{d} and O 2\textit{p} orbitals are found to be comparable, which shows the bonding nature of these orbitals. These comparable states are responsible for the covalent nature of bonding, due to which the compound also possesses covalent character. 
Further from the Table.1, it can be observed that the bond angle between O-Co-O is not 90$^\circ$. Hence, overlapping of orbitals is not as per ideal structure. Therefore, this can be a possible reason for the absence of states in the energy window -4.95 to -4.70 eV. In the region II, maximum states of Co 3\textit{d} with negligible states of O 2\textit{p} are found. Additionally, no states are found in the energy window -1.7 to -1.3 and 0.27 to 0.5. This dominating nature of Co 3\textit{d} states reflect its non-bonding character throughout this region. The states of O 2\textit{p} in region III are found slightly higher as compared to region II, which indicates the anti-bonding nature of Co 3\textit{d} and O 2\textit{p} orbitals. Thus, the idea of covalent nature of the compound can be extract from the above discussion. 

Fig.4(b) represents the PDOS obtained from the SP DFT calculation. The states are found to be asymmetrical for up and dn channels. From the figure, it can be observed that the shifting in states are found for up and dn channel with respect to Fig.4(a), in which the shifting in up channel is approximately twice with respect to dn channel. The observed asymmetrical nature of the PDOS depicts the magnetic nature of the compound. This is also reflected from the ground state energy obtained from SP and SUP calculation, where the SP ground state energy is $\sim$11 meV lower than the SUP energy. 
The calculated magnetic moment for the compound is found to be $\sim$1 $\mu$$_B$/f.u. In the region I, the asymmetrical nature of O 2\textit{p} states is an indicator of induced magnetism in O due to the strong hybridisation between the Co 3\textit{d} and O 2\textit{p} orbitals. Furthermore, the contributions of Co 3\textit{d} orbitals are found only in dn channel at E$_F$, which confirms the half-metallic nature of the compound.  
Moreover, the Co 3\textit{d} orbitals are dominating in regions II and III similar to SUP DFT PDOS  
and no states are found in the energy window 0 to 1 eV for the up channel and 0.44 to 1.25 eV for the dn channel. Thus from the above analysis, the understanding for magnetic nature of the compound can be made. 
 
NCO is a well known strongly correlated electron system due to the presence of partially filled Co 3\textit{d} orbitals \cite{tang}. We are aware that, the DFT alone can not treat properly the strongly correlated electron system \cite{anisim}. Therefore, to achieve more compatible results, we need to go beyond DFT. Thus, DFT+\textit{U} has been used for the calculation to deal with strongly correlated Co 3\textit{d} orbitals. The reported \textit{U} values for the Co 3\textit{d} are found in the range of 2.75-6 eV \cite{assadi,singh2017,sk2020,ma,singh2017un,xu,wissgott} for the study of electronic and transport properties of Co based compound. 
In the work of Singh \textit{et al}., Wissgott \textit et al., Shamim \textit et al., the reported value of \textit{U} for the S are 2.75, 3.5 and 4 eV, respectively \cite{singh2017un,wissgott,sk2020}. 
Therefore, we have performed DFT+\textit{U} calculation by considering \textit{U} in the range of 2-5 eV. From the calculation, it is found that \textit{U} = 4 eV giving the best match for our calculated S with the experimental value in the studied temperature range. Therefore, we have calculated all the properties by taking \textit{U} = 4 eV. Fig.4(c) shows the SP PDOS within DFT+\textit{U}(\textit{U} = 4 eV).
In region I, the comparable states of Co 3\textit{d} and O 2\textit{p} can be seen, which is similar to SP DFT PDOS. The decrement in states is found in region II, which causes a vanishing of the gap in energy range -1.7 eV to -1.30 eV. The gap of the up channel in regions II and III is also increased from $\sim$1.00 eV in SP DFT to $\sim$2 eV in SP DFT+\textit{U}. The dn channel states around the E$_F$ in case of SP DFT is sharply changing with energy, whereas, in DFT+\textit{U}, it is found approximately constant with change in energy. Generally, inclusion of \textit{U} in the strongly correlated electron system widens the gap between occupied and unoccupied states and also transfers the spectral weight, which can be the possible reason for the above phenomenon. 
Further, no states are found in the energy window $\sim$0.85-1.80 eV in dn channel of the region III. From the above, it can be observed that the Co 3\textit{d} states are largest occupied states around E$_F$ followed by insignificant states of O 2\textit{p}. It is well known that the states around E$_F$ will exclusively contribute in transport properties. 

\begin{figure}
\includegraphics[width=7.9cm, height=8.7cm]{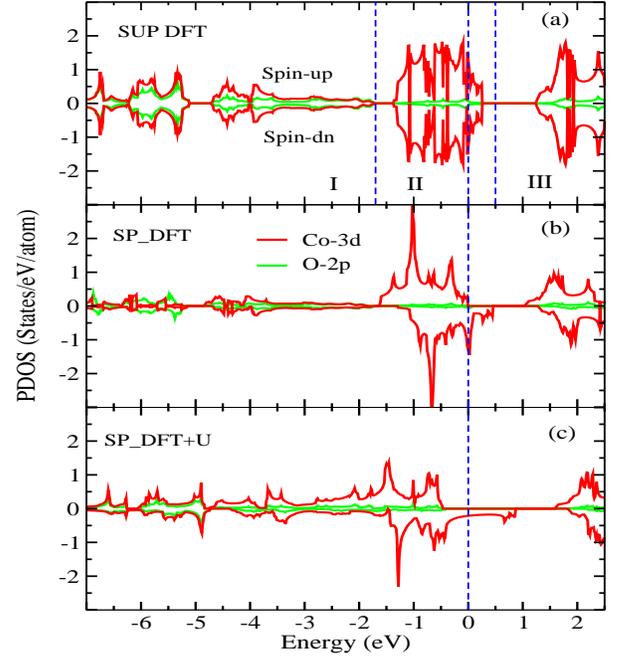}
\caption{Partial density of state(PDOS) for NaCo$_2$O$_4$ (a) Non-magnetic PDOS within DFT  (b) Magnetic PDOS obtained from DFT (c) Magnetic PDOS within DFT+\textit{U}(=4 eV).} 
\end{figure}

\subsection {Band structure} 
The calculation of SP electronic band dispersion has been performed for NCO within the DFT and DFT+\textit{U}(= 4 eV). Fig.5(a) and 5(b) show the dispersion curve in the DFT and DFT+\textit{U}. The dashed line at 0 eV represents the E$_F$. From the figure, it is noted that only two bands of the dn-spin channel are crossing the E$_F$ in each case. DFT bands are crossing at 9 k-points, while in DFT+\textit{U} these bands are crossing at 4 k-points. The crossing of these bands confirms the half metallic nature, which was also observed in PDOS. We have seen in the PDOS that around the E$_F$ the maximum states are from Co 3\textit{d}. On the basis of this, it is said that the crossing bands correspond to Co 3\textit{d} orbitals.
Further, the band width observed in DFT is $\sim$0.77 (0.7) eV for band 1 (2) which increases to $\sim$1.30 (1.37) eV for band 1 (2) after inclusion of \textit{U}. The bands of spin-up and spin-dn channels of DFT+\textit{U} are shifted away from the E$_F$ with respect to DFT. There is an indirect gap of $\sim$1.03 eV in spin-up channel in DFT, which is increased to $\sim$2.05 eV in DFT+\textit{U}. Due to the gap of spin-up channel, it can be expected that, as temperature increases there may be possibility of contribution of up channel states in the S also. However, it can be neglected due to the dominant behaviour of electrical conductivity of the dn channel in the two current model of the S which are discussed later in the manuscript. Hence, it is confirmed that, the transport properties have contribution from Co 3\textit{d} states of spin-dn channel only. 
 

\begin{figure}
\includegraphics[width=8.5cm, height=6.2cm]{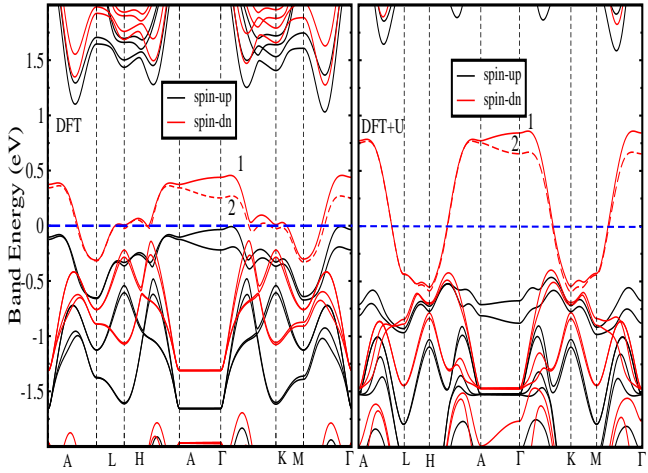}
\caption{DFT and DFT+\textit{U}(=4 eV) Band structure of NaCo$_2$O$_4$ in high symmetric directions.} 
\end{figure}

\subsection{Calculated transport properties}
\subsubsection{Seebeck coefficient}
In order to understand the experimentally obtained S on electronic level, we have calculated the transport properties of NCO within the DFT and DFT+\textit{U}(= 4 eV) with the help of BoltzTraP package \cite{madsen}. Here, constant relaxation time is considered for the BoltzTraP calculation. The two current model is used for the calculation of properties which includes both spin-up(dn) channels, respectively \cite{xiang}. Eq.3 shows the expression of the S used in the two current model

\begin{equation}
S  =\frac{S_\uparrow \sigma_\uparrow+S_\downarrow \sigma_\downarrow}{\sigma_\uparrow+ \sigma_\downarrow }
\end{equation} 
where, S$_\uparrow$$_\downarrow$ and $\sigma$$_\uparrow$$_\downarrow$ are Seebeck coefficient and electrical conductivity from up (dn) channel, respectively. 
\begin{table*}\label{tabb1}
\caption{\label{tabb}
\small{The calculated Seebeck coefficient ($\mu$V/K) corresponding to $\mu$=0 meV at different temperature.}} 
\begin{tabular}{lccccc}

\colrule
   
\colrule
\textrm{{Method}}
&\textrm{{S at 300 K }}
&\textrm{{S at 400 K }}
&\textrm{{S at 500 K}}
&\textrm{{S at 600 K}}
 \\

\colrule
Experimental  & 55 & 79  & 94 & 104 \\
DFT   & 12 & 15  &   18  &   20  &   \\       
DFT+U(=2 eV)  &  -1 &   -3  &   -8 &   -14 &\\
DFT+U(=3 eV)    &   -4 &   -3.50 &   -2 &   -0.40  & \\
DFT+U(=4 eV)  & -7 &   -9 &   -11 &   -13   &  \\
DFT+U(=5 eV)  & -7.50 &   -10 &   -12.50 &   -14.90   & \\
\colrule

\\
        
\colrule

\end{tabular}

\end{table*}

We have performed calculations for the S with respect to $\mu$ at different temperature within DFT and DFT+\textit{U}(=  2-5 eV). The calculated S corresponding to $\mu$ = 0 meV at various temperatures is shown in Table II, along with the experimental S. From the table, it is observed that the calculated value of S at $\mu$ = 0 meV is significantly deviated from the experimental S in the given temperature region. The sign of S in DFT is as per the experimental S, whereas, in the DFT+\textit{U}, it is opposite to the experimental in the whole studied temperature. The calculated value of S at $\mu$ = 0 meV is as per stoichiometry of the compound. However, it is difficult to synthesized the oxygen based compound with proper stoichiometry of oxygen. In addition to this, there is a presence of point defects at finite temperature, which increases with an increase in temperature. Due to these factors, the carrier density of synthesized sample deviated with the ideal stoichiometry compound, which changes the $\mu$ also. Therefore, the calculated result obtained at $\mu$ = 0 meV is not suitable to understand the experimental result on the electronic level. Hence, we have to find a different $\mu$ other than $\mu$ = 0 meV which can explain our experimental S upto the maximum possible extent.
 
\begin{table*}\label{tabb1}
\caption{\label{tabb}
\small{The Experimental and the calculated Seebeck coefficient ($\mu$V/K)at different temperature along with the standard deviation (n is the total number of data point in overall temperature).}} 
\begin{tabular}{lccccc}

\colrule
   
\colrule
\textrm{{Method}}
&\textrm{{S at 300 K }}
&\textrm{{S at 400 K }}
&\textrm{{S at 500 K}}
&\textrm{{S at 600 K}}
&\textrm{{($\sum$$|$S$_{exp}$-S$_{cal}$$|$)/n}} \\

\colrule
Experimental    &     55   &     78    & 94   & 104 & \\
DFT   & 72 &  79 &   94  &   95  &   4.37\\       
DFT+U(=2 eV)  &  67 &   82  &   94 &   100 & 4.97\\
DFT+U(=3 eV)    &   69 &   82 &   94 &   99  & 3.99\\
DFT+U(=4 eV)  & 71 &   83 &   94 &   102   & 3.93 \\
DFT+U(=5 eV)  & 70 &   84 &   94 &   101   & 3.98 \\
\colrule
\\
        
\colrule

\end{tabular}
\end{table*}
In order to get the best chemical potential, the results obtained for the S from DFT and DFT+\textit{U} are analysed at different $\mu$ corresponding to 300, 400, 500 and 600 K, respectively. The magnitude of average of absolute deviation between the computational and experimental S is calculated from these results, which is presented in the Table III. From the table, it is observed that the absolute deviation in S for DFT is $\sim$4.37 $\mu$V/K at $\mu$ = $\sim$705 meV corresponding to 500 K whereas, the least absolute deviation (3.93 $\mu$V/K) for DFT+\textit{U} is obtained at $\mu$=$\sim$752 meV corresponding to \textit{U} = 4 eV for 500 K. Therefore, it is found that the DFT+\textit{U}( = 4 eV) is giving the least deviation from the experimental result. This value of \textit{U} = 4 eV is consistent with the Shamim \textit et al. for Na$_0.$$_7$$_4$CoO$_2$ in the calculation of transport properties \cite{sk2020}.

\begin{figure}
\includegraphics[width=9.3cm, height=7cm]{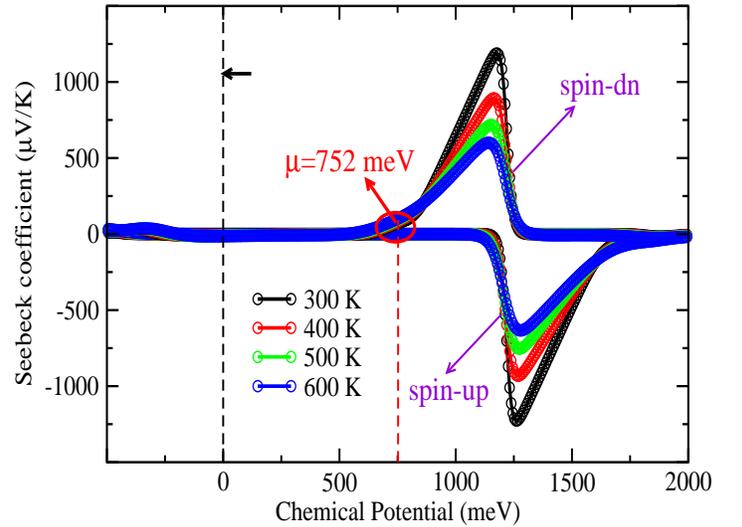} 
\caption{Variation of Seebeck coefficient with chemical potential of NaCo$_2$O$_4$ with in DFT+\textit{U}(=4 eV) corresponding to different temperature.} 
\end{figure}
We have obtained transport properties with respect to $\mu$ corresponding to various temperature for individual spin. The Fig.6 shows the variation of S with respect to $\mu$ at 300, 400, 500 and 600 K, respectively for spin-up (dn) channel. It can be observed from the figure, the contribution from up channel in S is negligible at $\mu$ = 752 meV as compared to dn channel. This is due to the insulating nature of the up channel which is mentioned above, as well as the extremely low electrical conductivity of this channel. The value of $\sigma$/$\tau$ around the above considered $\mu$ is in the range of $\sim$10$^1$$^0$ (10$^1$$^8$) $\Omega$$^-$$^1$cm$^-$$^1$s$^-$$^2$ for up (dn) channel at 500 K. Further, 
the obtained value of the S at the above noted $\mu$ is -4$\times$10$^-$$^6$ $\mu$V/K for the up channel, whereas it is in the range of $\sim$93 $\mu$V/K for the dn channel at 500 K. We have seen in the electronic properties that around the E$_F$ there are mainly Co 3\textit{d} states. Based on this it is said that the S is mainly contributed by Co 3\textit{d} dn channel states.
 
We have calculated the temperature dependent S for 300-600 within DFT and DFT+\textit{U} at $\mu$ equals to 705 and 752 meV and compared with the experimental S. Fig.7 shows the comparison of S for experimental, DFT and DFT+U. The S value at 300 K for experimental, DFT and DFT+\textit{U} is $\sim$55, $\sim$72 and $\sim$70 $\mu$V/K, respectively. Here, the deviation between experimental and DFT+\textit{U} S is larger in the initial temperature range, which decreases upto 500 K and then slightly increases from 500 to 600 K. Further, the S value at 600 K for experimental, DFT and DFT+\textit{U} is $\sim$104, $\sim$101 and $\sim$102 $\mu$V/K, respectively. Here, we can observed that the value of S obtained for this compound is very high as compared to other metals \cite{sk2022i,burkov}. These S values are showing the similar behaviour as of semiconducting material \cite{rashad,ahmad}. The strong electronic correlation effect can be the reason for this high S \cite{tang}, which made this as a different kind of compound other than general metal. Due to this property, this compound attracting much attention from the researchers.
Further, it is also observed that there is still slight deviation between experimental and calculated S at single chemical potential. The dependence of electronic structures on temperature and increase in defects with increase in temperature, which change the carrier concentration and ultimately chemical potential. These changes can be the reason for above deviation at single $\mu$ in S.

Moreover, the applicability of these S can be checked by the PF and corresponding ZT. In this respect, we have calculated the PF and maximum possible ZT by using the DFT+\textit{U} S, $\sigma$ and reported thermal conductivity \cite{ito} respectively. 

\begin{figure}
\includegraphics[width=9cm, height=6.5cm]{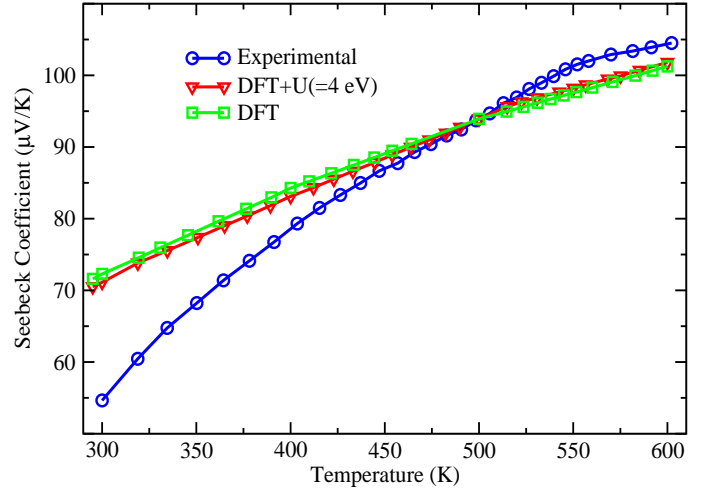} 
\caption{Comparison of experimental and calculated values of Seebeck coefficient as a function of temperature.} 
\end{figure}

\begin{figure*}[ht]
\includegraphics[width=0.8\linewidth, height=7.8cm]{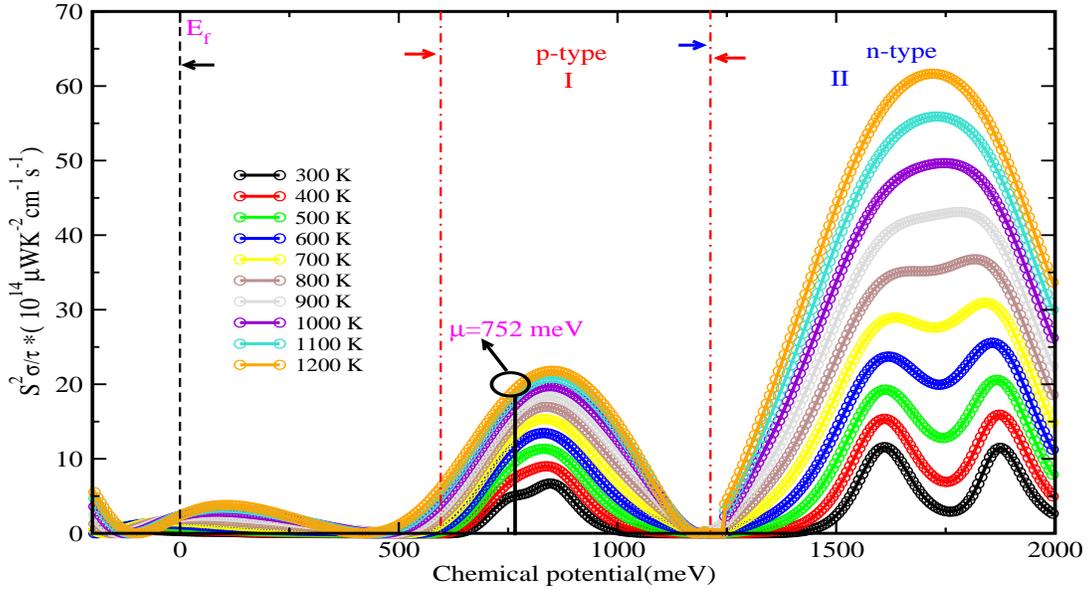} 
\caption{\small{Variation of Power Factor (PF) with respect to chemical potential corresponding to different temperature.}}
\end{figure*}
 \subsubsection{\textbf{Power factor, figure of merit }}
We have calculated the PF and ZT of NCO by using Eq.1 and 2. The PF is contributed by the S and $\sigma$ which can be seen in Eq.1. However, from the transport properties calculation by using the BoltzTraP package, we got the $\sigma$$/$$\tau$, where $\tau$ is constant. Therefore, we have plotted the S$^{2}$$\sigma$$/$$\tau$ in the Fig.8. The dashed line at 0 meV represents the E$_F$. We are mainly interested in the analysis of region I ($\sim$596 - 1215 meV) and region II ($\sim$1215 - 2000 meV) of this figure. Based on the analysis of calculated S (Fig.6), it is found that the region I(II) corresponds to p(n)-type doping behaviour. In the p-type region, we have found maximum S$^{2}$$\sigma$$/$$\tau$ for 500 K at $\sim$831 meV. However, it was seen in the earlier discussion of the manuscript that our calculated S was matched with the experimental S at $\mu$ = $\sim$752 meV at the same temperature. Therefore, this calculation of the S$^{2}$$\sigma$$/$$\tau$ suggests the possibility of significant amount of doping in the compound at the mentioned temperature. Further, there is also a maximum S$^{2}$$\sigma$$/$$\tau$ corresponding to p(n)-type at $\mu$ = $\sim$852 (1722) meV for 1200 K. These values are $\sim$22(61)$\times$10$^1$$^4$ $\mu$WK$^{-2}$cm$^{-1}$s$^{-1}$ for p(n) type of doping. These S$^{2}$$\sigma$$/$$\tau$ can enhance the ZT upto a certain limit, which improves the applicability of NCO for TE in higher temperatures. We have explored the applicability by calculating the ZT for the temperature upto 1200 K.

The calculation of ZT for both p(n)-types has been done by using Eq.2. The calculated values of S and $\sigma$ and the reported value of thermal conductivity (with small extrapolation upto 1200 K) \cite{ito} are used for the calculation. We have seen in the Fig.8, there is presence of $\tau$ also, and for the calculation of ZT this $\tau$ needs to be known. In this respect, we have calculated $\tau$ by comparing our calculated $\sigma$$/$$\tau$ with resistivity of Ito \textit{et al}. for 300 to $\sim$1200 (extrapolated from 1100-1200 K) K \cite{ito}. In order to compare the resistivity, we have matched the S of Ito \textit{et al}. with our calculated S at a individual temperature, and obtained the $\mu$ for $\sigma$$/$$\tau$. The value of $\tau$ obtained by comparing the resistivity at the particular $\mu$ is in the range of $\sim$1 to 0.415 $\times$$10^-$$^1$$^4$ s for 300 to 1200 K. In general, the value of $\tau$ at room temperature is in the range of $10^-$$^1$$^4$ to $10^-$$^1$$^5$ for metal \cite{ashcroft2011solid}, also Shamim \textit{et al}. have reported $\tau$ = $\sim$0.7 $\times$$10^-$$^1$$^4$ s at 300 K \cite{sk2020} for their metallic compound. Based on these values, it is said that the obtained value of $\tau$ can be used for the ZT calculation. 

The calculated value of $\tau$ is used for the calculation of ZT. Fig.9(a) shows the variation of ZT with respect to temperature in p-type range. We have presented ZT for the $\mu$ (= $\sim$752 meV) matched with our experimental result as well as for maximum p-type doping corresponding to individual temperature. The calculated value of ZT at $\mu$ = $\sim$752 meV is 0.08 (0.51) for 300 (1200) K, which is in well agreement with Ito \textit{et al}. and Nagira \textit{et al}. \cite{ito,nagira} for their compounds similar to NCO. Further, the value of ZT for p-type doping is higher as compared to the value obtained $\mu$ = $\sim$752 meV in overall temperature range. The ZT can be achieved upto $\sim$0.64 at 1200 K by careful doping of p-type element in NCO. In the work of Ito \textit{et al}., the experimental obtained value of ZT is $\sim$0.5 around 960 K \cite{ito}, which supports our calculated ZT. By getting these values of ZT, the applicability of the compound can be enhanced upto a certain limit. 

We have discussed the possibility of enhancing the ZT by p-type doping. However, in the TE generation (TEG), there is a requirement of n-type materials also. Therefore, we have to explore the possibilities of n-type materials in TEG by analysing the S$^{2}$$\sigma$$/$$\tau$ values corresponding to n-type of doping. In this respect, the ZT is calculated for n-type doping upto 1200 K by using S$^{2}$$\sigma$$/$$\tau$ and other properties which was considered for p-type. We have taken the maximum value of S$^{2}$$\sigma$$/$$\tau$ of n-type doping from Fig.8 at individual temperatures to maximise the possible ZT. Fig.9(b) shows the calculated value of ZT for n-type doping. The ZT is continuously increasing with temperature. The calculated values at 300 and 1200 K are found to be $\sim$0.1 and $\sim$1.8, respectively. In the work of Shamim \textit{et al}., the calculated value of ZT is reported as 2.7 at 1200 K for Na$_0$$_.$$_7$$_4$CoO$_2$ by taking constant $\tau$ (calculated at 300) for the whole temperature range \cite{sk2020}. However, we have considered the $\tau$ for ZT calculation at individual temperature, which enhance the reliability of these values. 
Further, it is observed that after doping of the heavier element and inclusion of nano-SiC in the synthesis of NCO, the value of $\kappa$ is found to decrease as compared to the value of $\kappa$ for NCO \cite{kurosaki,seetawan2,zhang2019}. This decrement in the $\kappa$ may lead to an increment in the ZT. Thus, if one can synthesized this compound by suitable n-type doping, one may get ZT = $\sim$1.8 or beyond. These higher ZT values of this compound deserved more attention for TE applications. 
\begin{figure}[ht]
\includegraphics[width=7.9cm, height=7.7cm]{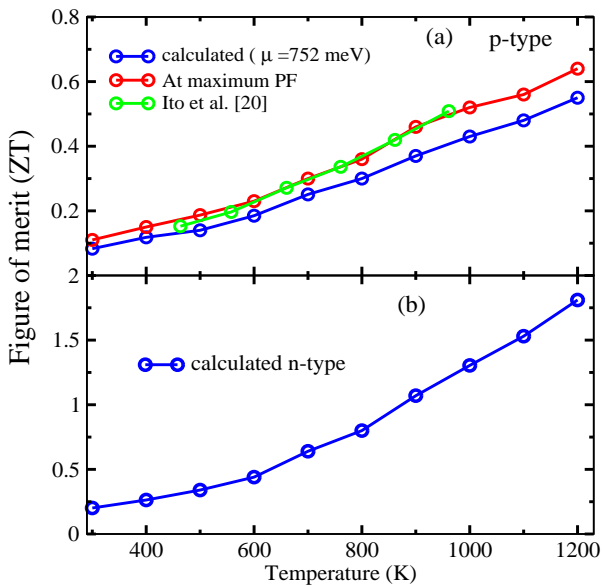} 
\caption{Variation of calculated ZT with temperature (a) at $\mu$ = $\sim$752 meV and for maximum of p-type doping (b) for n-type doping}
\end{figure}
Moreover, in the determination of transport properties, the available carrier concentration within the materials plays an important role. Therefore, these results became more reliable if the obtained carrier concentration is reasonable. Thus, in order to more verification or greater reliability of the above results, we have calculated carrier concentration at 100 K and compared with kitawaki \textit{et al} \cite{kitawaki}. The magnitude of S in the work of kitawaki \textit{et al}. is found to be 35$\mu$V/K and corresponding carrier concentration is 1$\times$10$^2$$^2$ cm$^-$$^3$ at 100 K \cite{kitawaki}. In our computational result the S value of 35$\mu$V/K is found at $\sim$744 meV and corresponding carrier concentration is in the range of $\sim$1.13$\times$10$^2$$^2$ cm$^-$$^3$ at 100 K. In addition to this, the calculated carrier concentration at 1200 K is 1.17$\times$10$^2$$^2$ (1.6$\times$10$^2$$^2$) cm$^-$$^3$ for p(n) type. The calculated carrier concentration in our work is in well agreement with the Kitawaki \textit{et al}. at the same temperature. Above argument indicates the reliability of the predicted ZT value. 
\subsubsection{\textbf{Conclusion}}
Firstly, we have studied temperature dependent Seebeck coefficient (S) for the NaCo$_2$O$_4$ by using combined experimental and computational approach. 
The positive and monotonic increasing nature of S is found within the temperature range of 300-600 K, which shows the dominant p-type behaviour of the compound.
The experimental value of S is found to be $\sim$55 (103) $\mu$v/K at 300 (600) K. In addition to this, the electronic structures (PDOS and Band dispersion curves) and electronic transport properties (S, power factor and ZT) have been studied by using DFT+\textit{U} (\textit{U} = 4 eV). These electronic structures and electronic transport properties explain the experimental S in a better way. The magnetic and half metallic nature is confirmed by the spin polarised PDOS. The temperature and chemical potential ($\mu$) dependent S$^{2}$$\sigma$$/$$\tau$ is calculated by using the electronic transport properties. The maximum S$^{2}$$\sigma$$/$$\tau$ is found at $\mu$ equals to $\sim$852 and 1722 meV for p and n type doping and respective carrier concentration is obtained as 1.17 $\times$10$^2$$^2$ and 1.6$\times$10$^2$$^2$ cm$^-$$^3$ at 1200 K. By using S$^{2}$$\sigma$$/$$\tau$, the ZT has been calculated for p and n-type within temperature range of 300-1200 K. The maximum ZT obtained for p (n)-type doping is 0.64 (1.8) at 1200 K. The calculated value of ZT suggests that, the n-doped NaCo$_2$O$_4$ has significant potential for high temperature thermoelectric applications.
\bibliography{ref}
\bibliographystyle{apsrev4-1}

\end{document}